# A Survey on Machine Learning for Geo-Distributed Cloud Data Center Management

Ninad Hogade, *Student Member, IEEE,* and Sudeep Pasricha, *Senior Member, IEEE*

**Abstract** — Cloud workloads today are typically managed in a distributed environment and processed across geographically distributed data centers. Cloud service providers have been distributing data centers globally to reduce operating costs while also improving quality of service by using intelligent workload and resource management strategies. Such large scale and complex orchestration of software workload and hardware resources remains a difficult problem to solve efficiently. Researchers and practitioners have been trying to address this problem by proposing a variety of cloud management techniques. Mathematical optimization techniques have historically been used to address cloud management issues. But these techniques are difficult to scale to geo-distributed problem sizes and have limited applicability in dynamic heterogeneous system environments, forcing cloud service providers to explore intelligent data-driven and Machine Learning (ML) based alternatives. The characterization, prediction, control, and optimization of complex, heterogeneous, and ever-changing distributed cloud resources and workloads employing ML methodologies have received much attention in recent years. In this article, we review the state-of-the-art ML techniques for the cloud data center management problem. We examine the challenges and the issues in current research focused on ML for cloud management and explore strategies for addressing these issues. We also discuss advantages and disadvantages of ML techniques presented in the recent literature and make recommendations for future research directions.

——————————— ◆ ———————————

## 1 INTRODUCTION

In October 2021, 4.88 billion people across the globe were connected to the internet. That is nearly 62% of the global population, and this number is growing at a rate of 4.8% per year [1]. Most people use smartphones to connect to the Internet, but the use of portable computers, wearable devices, smart home and security products, Internet-of-Things (IoT) devices, etc., is continuously increasing. To handle the resulting massive volumes of internet data, companies of all sizes are looking for ways to achieve reliable and secure storage and data processing, with easy administration of internal and external data. This can be achieved with the help of cloud computing. Cloud computing is the delivery of virtual resources (servers, storage, apps, online services, etc.) over the Internet by centralized systems situated far away from end users. For simplicity in this article, we refer to all hardware resources as "resources." The trend of moving computation, data storage, and applications into cloud data centers has facilitated ubiquitous access to shared computing and storage resources on demand. Among their many benefits, cloud computing services have helped improve IT workflows, reduce costs by removing on-premises servers, increase scalability, ease maintenance, improve deployment speed, lessen downtime, enhance security, and facilitate mobility and flexibility of work practices.

The ever-growing use of the internet, data processing, storage, and advancement in modern technology has led to an increase in the use of cloud computing over time. The worldwide cloud computing market is projected to grow from $445.3 billion in 2021 to $947.3 billion by 2026 [2]. Because of the COVID-19 pandemic, digital business transformation has become more difficult and urgent. As a result of the pandemic's economic impact, major companies are providing clients with cost-effective digital solutions. The need for cloud services has surged as a result of the sudden closures of offices, schools, and businesses due to the pandemic. By 2025, Gartner [3] predicts that over 85% of businesses will have adopted the cloud-first strategy, with over 95% of new workloads being deployed on cloud-native platforms (up from 30% in 2021). Cloud revenue will overtake non-cloud income in IT markets during the next few years.

### 1.1 Need for Geo-distributed Cloud Data Centers

Cloud service providers are increasingly developing additional data centers to fuel the cloud computing boom, bolstering the capabilities of their cloud computing offerings. While centralized data centers were once commonplace, there has been a trend in recent years toward dispersing data centers throughout different geographical regions [4], [5]. There are many advantages to distribute data centers globally. It brings businesses closer to their clients by improving performance (e.g., latency) and lowering network expenses. Due to the redundancy that distributed data centers enable, it also provides superior resilience to unanticipated failures (e.g., environmental disasters). It enables businesses to handle data with various criteria as required by a specific country/region, industry, or workflow. It also aids in the compliance of regional data storage and privacy laws because some data or information is not allowed to be disseminated or processed outside of a specific geographic region. Corporations can utilize the distributed cloud to enforce restrictions while still adhering to data privacy regulations. Another compelling reason to geographically distribute data centers is to reduce energy costs by taking advantage of electricity pricing and renewable energy differences across different regions. Because of the aforementioned reasons and the increasing trend of distributing data centers globally, in this article, we review research efforts and recent developments that focus on "geo-distributed" cloud data centers.

• N. Hogade and S. Pasricha are with the Dept. of Electrical and Computer Engineering, Colorado State University, Fort Collins, CO 80523. E-mail: {ninad.hogade, sudeep}@colostate.edu.

## 1.2 Challenges with Geo-distributed Cloud Services

Key cloud service providers such as Amazon Web Services (AWS), Microsoft Azure, Google Cloud, and Alibaba Cloud run massive data centers around the world. All of these data centers are packed with high-core-count CPUs, terabytes of RAM, and petabytes of storage. Bloomberg's latest report [6] shows that data centers currently account for roughly 1% of worldwide electricity usage, and that percentage is expected to climb to 3 - 8% in the next decade. The amount of energy used by data centers around the world has doubled in the last decade, and some studies predict that it may triple or even quadruple in the next decade [7]. The annual electricity expenditure for powering data centers is continually increasing, so minimizing such expenditures is critical. As an example, by 2023, China's data centers are expected to consume more energy than the entire country of Australia [8]. Such annual energy expenditures sometimes exceed the cost of purchasing data center equipment. Thus, determining how to manage cloud computing components in a cost-effective, carbon/energy-saving, and balanced manner has become a significant focus of researchers working in the cloud computing area.

Another challenge is in providing timely access to cloud services, without incurring SLA and QoS violations. A Service-level Agreement (SLA) is a contract between a service provider and a client that defines the scope of the service, e.g., service quality, availability, and vendor responsibilities. From defining services to the end of the agreement, SLAs normally have a lot of components and are also defined at many levels. Quality of service (QoS) is the description of a service's overall performance, such as a computer network or a cloud computing service, as observed by its users. Many network service components such as goodput, availability, latency, transmission delay, out-of-order delivery, errors, and packet loss, are frequently used to assess QoS for cloud services. High-volume data collection and processing in cloud data centers often leads to high latencies and network bandwidth use. According to an Amazon analysis, a 100ms increase in web page loading time cost the company 1% of its sales. Google discovered that adding 0.5 seconds to the time it takes to generate a search query result reduced traffic to its site by 20% [9]. For interactive and time-critical applications, lower latencies can greatly improve the quality of the experience. This highlights the importance of preventing SLA and QoS violations.

Cloud management approaches are often employed to reduce the energy consumption, carbon emissions, network latency, query completion time, data center operating costs, and QoS and SLA violations. However, orchestrating the sophisticated, heterogeneous, large-scale software and hardware systems in complex network settings, and automating computing platforms involved in geo-distributed cloud computing systems, is a very difficult problem. Researchers in this field have been employing heuristics and algorithms for sub-problems such as low overhead virtual machine (VM) migration, queuing minimization, latency-aware service/workload migration, and performance/energy optimal resource allocation. Current mathematical methods to resolve such problems employ many strategies, such as heuristic methods, meta-heuristic algorithms, probability algorithms, hybrid algorithms, and dynamic programming methods. These mathematical methods have limited relevance in large scale dynamic HPC systems and are difficult to scale to geo-distributed levels. As a result, cloud service providers have been obliged to investigate intelligent data-driven and Machine Learning based alternatives.

## 2 OVERVIEW OF MACHINE LEARNING METHODS

Machine Learning (ML) refers to a machine's ability to learn from data. Without being explicitly programmed, ML allows computers to learn how to perform tasks, such as predicting outcomes and classifying objects. One of its key assumptions is that by utilizing training data and statistical approaches, it is possible to develop algorithms that can anticipate future or previously unseen values. ML has slowly established itself as the de-facto solution in a variety of enterprise applications, including speech recognition, self-driving cars, business intelligence, web search, fraud detection, purchase recommendations, and customer service, to name only a few examples. The availability of enormous datasets and continued advances in both ML theory and the computational capabilities of servers are largely responsible for this accomplishment. In recent years, big tech companies have made considerable investments in ML and Artificial Intelligence (AI) by introducing new services and undertaking major reorganizations to strategically place AI in their organizational structures [10].

There are three broad ML paradigms in widespread use today: supervised learning, unsupervised learning, and Reinforcement Learning (RL). The purpose of supervised learning is to find a function that best approximates the relationship between inputs and outputs, given a sample of input data and desired outputs. Unsupervised learning, on the other hand, has no labeled outputs and instead seeks to infer the inherent structure existing in a set of input data points. RL is an ML algorithm that rewards desirable behaviors while penalizing undesirable ones. We briefly introduce the ML algorithms included in this survey in the following sub-sections.

### 2.1 Clustering

Clustering is the process of dividing a set of data points into groups so that data points in the same group are more like each other and different than the data points in other groups. Clustering methods are unsupervised ML algorithms. The goal of *k*-means clustering, a widely used clustering algorithm, is to divide *n* data points into *k* clusters, with each data point belonging to the cluster with the closest mean (cluster centroid or cluster center), which serves as the cluster's prototype. Spectral clustering is another method that has its origins in graph theory and is used to discover clusters of nodes in a graph based on the edges that connect them.

### 2.2 Regression

A problem is called regression when the purpose is to forecast a continuous or quantitative output value. Regression is a subfield of supervised ML. A linear model, such as one that predicts a linear relationship between the input variables $x$ and a single output variable $y$, is known as *Linear Regression (LR)*. The link between the input variable $x$ and the output variable $y$ is treated as an *n*th degree polynomial in $x$ in *Polynomial LR (Poly LR)*. The *Least Absolute Shrinkage and Selection Operator (LASSO)* is a regression analysis approach that does both variable selection and regularization to improve the statistical model's prediction accuracy and interpretability. In *Decision Trees (DTs)*, the goal is to learn simple decision rules from data attributes to develop a model that predicts the value of a target variable. *Random forest (RF)* algorithms use ensemble learning method for regression. The ensemble learning method combines predictions from several ML algorithms to get a more accurate forecast than a single model. RF regression works by training a large number of DTs. Gradient boosting is a regression technique that generates a prediction model from an ensemble of weak prediction models, most commonly DTs. The resulting technique is called *Gradient-Boosted Regression Trees (GBRTs)* when a DT is the weak learner. *eXtreme Gradient Boosting (XGBoost)* is an efficient and popular open-source implementation of the gradient boosted trees. The engineering goal of XGBoost is to push the computational resources' limits for boosted tree algorithms. The *k-Nearest Neighbors (k-NN)* regression technique approximates the relationship between independent variables and continuous outcomes by averaging observations from the same neighborhood. *Support Vector Machine (SVM)* regression is a supervised learning method for predicting discrete values. SVM regression's key concept is to identify the best-fit line. The hyperplane with the most points is the best fit line in this case. *Auto-Regressive Integrated Moving Average (ARIMA)* is a generalization of the simpler Auto-Regressive Moving Average (ARMA) and adds the notion of integration. Both models are used to fit time series data to better understand or anticipate future points in the series. *Seasonal ARIMA (SARIMA)* is an ARIMA variant that supports univariate time series data with an explicit seasonal component.

## 2.3 Neural Networks

*Neural networks (NNs)*, also known as artificial NNs (ANNs), are a class of ML algorithms whose name and structure are derived from the human brain. These networks resemble networks of biological neurons and mimic the manner in which they communicate with one another. NNs are made up of node layers, consisting of an input layer, one or more hidden layers, and an output layer. Each node, or artificial neuron, is connected to the others and has a weight and threshold associated with it. If a node's output exceeds a certain threshold value, the node is activated, and data is sent to the next layer of the network. Otherwise, no data is sent on to the network's next layer. To learn and increase their accuracy over time, NNs depend on training data. However, once these learning algorithms have been fine-tuned for accuracy, they become strong tools in area of computer science and AI, allowing us to rapidly cluster, classify, or predict data. When compared to manual identification by human analysts, tasks in the domains of voice or image recognition can take seconds with NNs rather than hours.

*Stochastic configuration networks (SCNs)* use a supervised learning mechanism to automatically and rapidly construct universal approximators that achieve promising performance for solving regression problems. A *Multi-Layer Perceptron (MLP)* is a type of feedforward NN. A perceptron is a model of a single neuron that served as a predecessor to larger NNs. *Deep Learning (DL)* is a subset of ML, which is essentially a NN that has three or more layers. While a single-layer NN may produce approximate predictions, extra hidden layers can help to optimize and enhance accuracy. A *Convolutional Neural Network (ConvNet/CNN)* is a DL model that can take an input image, assign relevance (learnable weights and biases) to numerous objects in the image, and distinguish one from the other. A *Recurrent Neural Network (RNN)* is an NN that works with time series or sequential data. RNNs have hidden states and allow previous outputs to be utilized as inputs. Language translation, image captioning, natural language processing (NLP), and speech recognition are just a few of the challenges that these RNNs are widely utilized for. *Long short-term memory (LSTM)* is a type of RNN with a more complex neuron structure. LSTMs can handle predictions with individual data inputs (such as images) as well as long data sequences (such as speech or video). LSTM cells uses a gating mechanism with three gates: input, output, and forget. *Gated recurrent units (GRUs)* are similar to LSTMs but have only two gates: reset and update. Because GRUs have fewer gates than LSTMs, they are less complicated. If the dataset is small, GRUs are generally preferable; otherwise, for larger datasets, LSTMs often demonstrate better performance.

## 2.4 Reinforcement Learning

Reinforcement learning (RL) is an ML technique that allows an agent to learn by trial and error in an interactive environment using feedback from its own actions and experiences. The *agent* represents the RL algorithm, while the *environment* refers to the item on which the agent is acting. The environment sends a state to the agent, which then takes action in response to that condition based on its knowledge. The environment then sends the agent a pair of next state and reward information. To evaluate its latest action, the agent will update its knowledge using the reward returned by the environment. The cycle continues until the environment sends a terminal state, bringing the episode to an end. Below are the few important terms used in RL.

- *Action:* All the moves available to the agent.
- *State:* Current state returned by the environment.
- *Reward:* An immediate return (feedback) from the environment to assess the quality of the previous action.
- *Policy:* The agent's strategy for determining the next course of action based on the current state.
- *Value:* The predicted long-term return with discount, in contrast to the short-term reward.
- *Q-value* or *action-value:* The long-term return of the current state, taking an action, under a policy.
- *Model:* The agent's perspective on the environment, which converts state-action pairings into probability distributions over states.

It is worth noting that not every RL agent makes use of a model of its surroundings. *Model-based* RL tries to model the environment and then determine the best policy based on the model it has learned. On the other hand, in *model-free RL*, the agent uses trial and error experience to output the optimal policy. Policy optimization and Q-learning are the two basic methodologies for representing agents using model-free reinforcement learning. In *policy optimization* or policy-iteration methods, the policy function that maps states to actions is directly learned by the agent. A value function is not used to decide the policy. In *Policy Gradient (PG)*, gradient descent is used to optimize parametrized policies with respect to the expected return (long-term cumulative reward). *Actor-critic* is another policy optimization method where the critic estimates the action-value (Q-value) or state-value function and the actor updates the policy distribution in the direction suggested by the critic. The *Q-learning* algorithm attempts to determine the best action given the current situation using a lookup table called Q-table, where the maximum expected reward for actions at each state are stored. *Deep Q Neural Network (DQN)* combines Q-learning with NNs. Instead of building Q-table, NNs approximate Q-values for each action based on the state. *Deep Deterministic Policy Gradient (DDPG)* is an algorithm that learns both a Q-function and a policy at the same time. It learns the Q-function with off-policy data and then utilizes the Q-function to learn the policy. *Deep reinforcement learning (DRL)* is a combination of reinforcement learning and deep learning. Deep learning is included into DRL, allowing agents to make decisions based on unstructured input data without having to manually build the state space. DRL algorithms can process enormous amounts of input and determine what actions to take to optimize an objective. The use of RL in a multi-agent system is known as *Multi-Agent Reinforcement Learning (MARL)*. Typically, agents develop their decisions as a result of their previous experiences. Here, an agent, in particular, must understand how to coordinate with other agents similar to game theory.

## 3 SURVEY ORGANIZATION

The purpose of this article is to discuss the applications of ML in geo-distributed cloud data center management, as well as the challenges that remain unsolved, to highlight opportunities for future work in this area. We do not limit our scope to a specific context, scenario, or eco-system in this regard, but rather focus on a variety of techniques proposed to address geo-distributed data center management; and discuss how these techniques are influencing the future generation of cloud computing architectures. Such architectures are highly heterogeneous and complex, to the point that a completely engineered approach to modeling and prediction could be needlessly complicated, if not outright futile. In such environments, cloud optimization plays a very important role. The process of correctly identifying and assigning the right resources to workloads (tasks or applications) or vice versa is known as cloud optimization. We specifically focus on techniques that use ML for prediction, classification, profiling of resources and workloads, application/task placement, migration, cloud optimization, and hybrid ML methods across the spectrum of geo-distributed computing scenarios. Accordingly, this article deconstructs and analyzes the challenge of reliable geo-distributed cloud data center management before surveying ML based approaches that solve this problem (or parts of it). The contributions of this article can be summarized as follows:

- we decompose the geo-distributed cloud data center management problem into various sub-problems;
- our study comprehensively reviews state-of-the-art ML-based techniques for workload and resource profiling, parameter prediction, and cloud optimization;
- we discuss challenges and future directions for ML in geo-distributed cloud data center management.

We organize the rest of the article as follows. In Section 4, we review relevant prior surveys on themes related to the ones we focus on. We define the responsibilities of cloud managers and provide a classification of sub-problems that form the structure of the survey in Section 5. Sections 6 surveys recent ML based methods for addressing various aspects of geo-distributed cloud data center management and compares the approaches. Section 7 presents an overview of the reviewed articles, identifies open issues, and discusses future opportunities. Lastly, Section 8 concludes the survey.

## 4 RELATED SURVEYS

### 4.1 Cloud Management

There are many previous review articles that have explored resource management, workload management, security, VM allocation, energy efficiency, and load balancing for cloud computing. Some recent reviews [11], [12], [13] presented a multi-dimensional classification of existing cloud resource management solutions based on their scheduling architectures, objectives, and methods. In [14], various clustering, optimization, and ML methods used in cloud resource management to improve energy efficiency and performance are compared, classified, and analyzed. In [15], [16], and [17] an overview is provided for existing cloud computing load balancing algorithms and their performance metrics. These works also discussed the advantages and disadvantages of the chosen load balancing algorithms. In [18] a study of various available resource allocation methods, load balancing techniques, scheduling techniques and admission control techniques for cloud computing is presented along with an analysis of the advantages and disadvantages of each method. Based on our analysis of these surveys, we observed that some reviews considered a few ML based studies, but they do not comprehensively review ML based cloud management that presents the importance of the ML techniques in cloud computing.

### 4.2 Machine Learning based Cloud Management

Several recent surveys have considered ML based cloud management. Among these, [19] divided the cloud resource scheduling problem into various objectives, such as time optimization, energy consumption optimization, and load balancing optimization. The authors discussed and researched the architecture of RL and Deep Reinforcement Learning (DRL). The review progressively evaluated various architectures of RL and DRL, offering a unified view for the subsequent RL or DRL in cloud computing resource scheduling. In [20], the authors looked at how the topic of reliable resource provisioning in joint edge-cloud systems has been addressed in the scientific literature, and what strategies have been utilized to increase the reliability of distributed applications in diverse and heterogeneous network environments. The survey discussed how the characterization, management, and control of complex distributed systems utilizing ML methodologies are given significant attention due to the problem's complexity. The survey is organized around a three-part deconstruction of the resource provisioning problem: workload characterization and prediction, component placement, system consolidation, and application elasticity and remediation. The authors in [21] divided cloud resource management into four major categories: resource provisioning, resource scheduling, resource allocation, and energy efficiency. After presenting a comparison between works in these categories, they suggested the most suitable ML model for each category. In [22], a broad survey of works that leverage ML techniques for resource management in cloud computing is presented. Unlike other works, this review categorized and analyzed the included studies on the basis of three types of ML models: supervised learning, unsupervised learning, and RL. This review further evaluated the ML techniques by the type of features they use, the outputs they produce, and the objectives they try to optimize. In [23] a detailed review of challenges in ML-based cloud computing resource management is presented. The survey also discussed state-of-the-art approaches to resolve these challenges, along with their pros and cons. Finally, the authors proposed possible future research directions based on identified limitations.

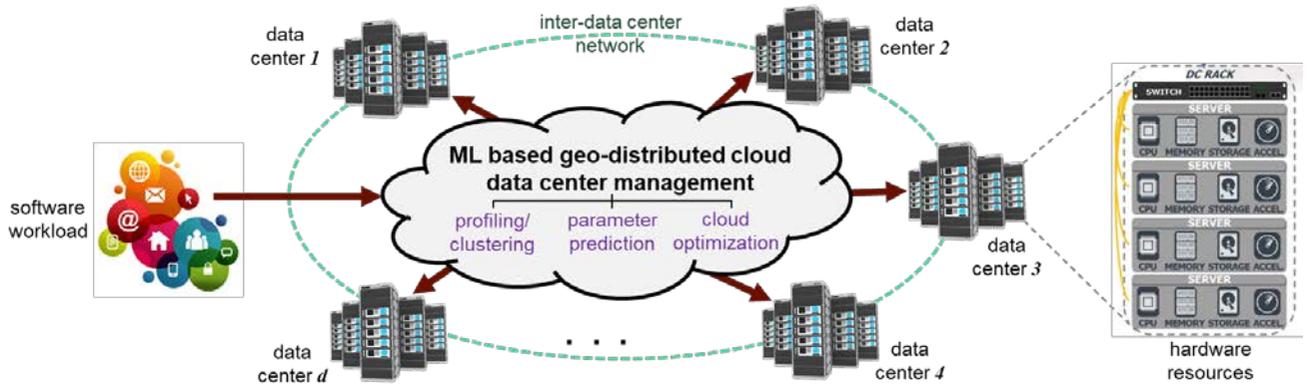

Fig. 1. Machine Learning (ML) based geo-distributed cloud data center management

### 4.3 Machine Learning based Geo-Distributed Cloud Data Center Management

While it is very useful to employ ML in distributed cloud computing systems and to assess the progress of the recent research and engineering of these systems, none of the recent articles go into detail about ML-based cloud management for geo-distributed data centers, nor do they go into the specifics of the challenges and issues that exist in the current state-of-the-art and future research directions on this theme. As discussed in our previous work [24], geo-distributed cloud data centers can exploit differences between electricity prices, net metering, peak demand pricing distribution, resource availability, data transfer/network prices, co-location in servers, and renewable energy at various data center locations. Additionally, geo-distributed data centers give higher resilience to unplanned failures due to their redundancy. As a result, we believe that it is the right time to now present a comprehensive survey that discusses various ML algorithms and their applications to different problem contexts within cloud management scenarios for geo-distributed data centers, as well as their shortcomings, challenges, and future directions, in accordance with our vision. Thus, before moving further with their new ideas in this regard, researchers can use this article to assess the current ML scenarios in geo-distributed cloud management and their limitations.

## 5 COMPONENTS OF CLOUD MANAGEMENT

Reliable cloud management for data centers is a complex problem, especially when it is examined in a geo-distributed cloud environment where distributed infrastructure is utilized to host numerous heterogeneous workloads coming from different geographical regions. Each workload typically has its own set of requirements, and there is a high possibility that controlling operations or tuning the performance of one workload that is using one or multiple resources will have some effect on the others e.g., workload co-location interference effects [25]. Additionally, the problem involves deployment and operation with workloads in geographically dispersed heterogeneous resource environments. In such complex environments, solving the problem completely and optimally is a difficult task, and viable solutions must involve a combination of various methods to achieve the predictability and controllability of both the (software) workload and (hardware) resource performance. This necessitates a thorough investigation of all parts of the problem to develop a holistic understanding of the problem and the challenges while developing a solution. For this purpose, as shown in Figure 1, we divide and discuss aspects of geo-distributed cloud management across three categories: profiling/clustering, parameter prediction, and cloud optimization.

### 5.1 Workload and Resource Profiling

Cloud workloads can be classified into different types based on various characteristics and perspectives, such as transactional or non-transactional, sequential or non-sequential/random, and memory intensity. In terms of both resource type and volume, the need for resources varies depending on the workload. A cloud application workload's components are designed to carry out certain activities that are likely to use various types of resources (e.g., CPUs or GPUs). Moreover, because of the wide geographic distribution of data centers over networks, workloads can be distributed in different ways across different data center locations. These characteristics lead to high complexity and challenges in workload and resource profiling. The ability to understand workload and resource behavior using ML appears to help improve an application's performance and reliability, by allowing cloud providers to ensure the availability of sufficient resources to support any future workloads. Therefore, many cloud management techniques require workload and resources to be profiled or clustered into a specific category, before the techniques can be effectively applied.

## 5.2 Parameter Prediction

In the age of pervasive edge and IoT computing, the workloads running across cloud environments are steadily increasing in both complexity and volume. The heterogeneity of workloads and the number of users using a given cloud application service at various times introduces changes in cloud workloads and the corresponding resources that the workloads need to use across data center locations. In other words, cloud applications experience variations of workload arrival and resource usage in a manner that can be difficult to predict. The ability to predict the spatio-temporal distribution of future workloads or their parameters in advance with ML brings benefits for geo-distributed cloud management solutions, i.e., the system gains the ability to scale proactively to satisfy real-time resource requirements. Or, in some cases, the workload can be allocated to resources effectively if future utilization states of the resources are known. In this way, it is possible to improve workload allocation and optimize resource utilization while guaranteeing QoS requirements and SLAs, which in turn enhances the efficacy and performance of geo-distributed cloud management.

## 5.3 Cloud Optimization

Cloud optimization techniques are typically implemented using either static or dynamic algorithms. While the workload placement or the resource allocation rules are predetermined in a static scheme, in a dynamic scheme the workload is dynamically processed and allocated to the resources. Even though static techniques are simple and usually more stable, they do not achieve optimal results with heterogeneous workloads and distributed resources subject to erratic workload distribution and unpredictable resource behavior. The cloud optimization techniques allocate or release resources on the fly according to the demand of the workload, thus reducing the operating costs while maintaining QoS requirements and SLAs. Ideally, cloud resources and workloads need to be dynamically configurable, programmable, and optimizable with the help of ML, without any human input. An important component of cloud resource management involves Virtualized Network Functions (VNFs) that are created on open computing platforms and run virtual network services on common servers in cloud data centers. Virtualized routers, firewalls, WAN optimization, and network address translation (NAT) services are all examples of VNFs. Virtual machines (VMs) running on common virtualization infrastructure software such as VMWare or Kernel-based Virtual Machine (KVM) run the majority of VNFs. Network service providers must configure network traffic according to flow-level granularity, lowering network costs and ensuring a good user experience. As workload and resource demands vary over time in cloud environments, it is also important to dynamically scale the VNF instances, with the help of ML.

## 6 SURVEY OF RELEVANT WORKS

ML and cloud computing have become popular and widely adopted in industry. A recent trend has been to explore how ML may help with cloud management for geo-distributed data centers. In the past few years there has been a continuous increase in the numbers of studies in this area. For this survey, we selected 22 articles related to the ML based geo-distributed cloud data center management. The articles have been classified into four main groups. A taxonomic classification of the literature surveyed in this review is shown in Figure 2. In the figure and the rest of the article, "resource" refers to hardware systems, whereas "workload" refers to software applications or tasks. The following subsections discuss the prior works in these four main groups in more detail.

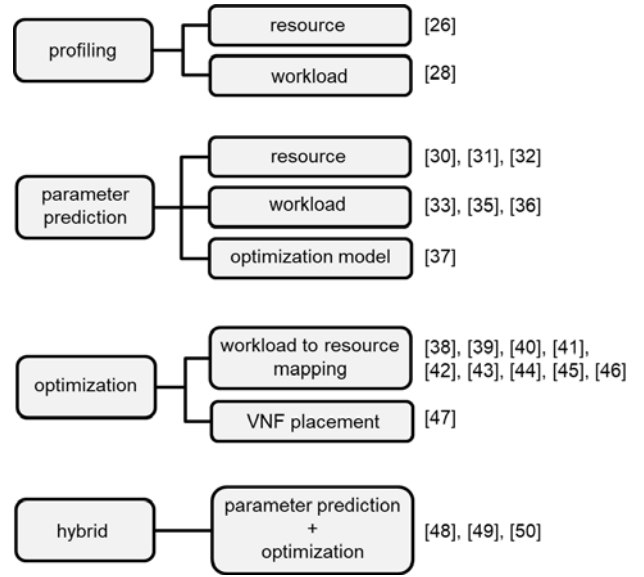

Fig. 2. Taxonomic classification of the literature surveyed

## 6.1 Profiling

### 6.1.1 Resource Profiling

The goal of [26] is to select the optimal data center (resources) among a set of geo-distributed data centers based on SLA standards and the needs of the user. The SLA here is comprised of a set of service-level objects (SLOs), but in this study, the four most essential ones are cost, response time, availability, and reliability. After receiving the user's request in a geo-distributed cloud environment, firstly, the data are normalized using the Z score normalization method to remove abnormal distribution of various SLOs. Then the number of clusters are determined according to four SLA criteria, where 3844 data centers are clustered using the $k$-means algorithm. In the next step, based on the longitude and latitude of the user's location, the nearest geographical cluster is selected. Furthermore, accessibility and reliability are maximized, while cost and response time are minimized, and the multi-objective NSGA-II [27] algorithm is used to select the best data center based on these SLOs.

### 6.1.2 Workload Profiling

Classical approaches for uniformly partitioning data across distributed nodes are the norm in many major distributed data storage solutions, such as Cassandra and Hadoop Distributed File System (HDFS). These methods can generate network congestion for data-intensive services, lowering system bandwidth. Data-intensive services, in contrast to MapReduce workloads, require access to many datasets throughout each session. To address these issues, the authors in [28] proposed a scalable method for distributing data-intensive services across geographically distributed cloud data centers. The authors first constructed a hypergraph using a given set of data items, and the set of user check-ins, each comprising a data request pattern being obtained from a data center location. Hypergraphs allow for the capturing of multi-way relationships, making it easier to model data item to data item, and data item to datacenter location relationships. The dataset used in the experiments is a trace of a location-based online social network – Gowalla, available publicly from the Stanford Network Analysis Platform (SNAP) repository [29]. Using spectral clustering on hypergraphs, the proposed approach divided a set of data items into geo-distributed cloud data center locations to minimize the weighted average of the per unit cost of outgoing traffic from each datacenter, the inter datacenter latency for each datacenter pair, the per unit storage cost of each datacenter, and the average number of datacenters accessed by the data items requested in each request pattern. Furthermore, the spectral clustering algorithm employed randomized techniques to create low-rank approximations of the hypergraph matrix, allowing for greater scalability in the computation of the hypergraph Laplacian spectra.

## 6.2 Prediction

### 6.2.1 Resource Prediction

In a Geographical Load Balancing (GLB) policy, the amount of load delivered to each local data center is determined by the overall system's conditions rather than the local situation. Regulation Service (RS) is a type of service provided by the power market to continuously balance supply and demand at a very granular level while also offering economic benefits to consumers and the network/utility. Optimizing RS for local data centers in a geo-distributed cloud is a difficult problem to solve. Motivated by this phenomenon, [30] used SARIMA to predict the incoming workload at a global level. Then the authors used a CNN-based ML model to forecast/output the power consumption of local data centers. Inputs to the models are the data center's outdoor temperature, solar power, electricity price, and workload arrival rate. The model used Rectified Linear Unit (ReLU) as the activation function, Adam optimizer, and Root Mean Squared Error (RMSE) as an evaluation metric. To develop a prediction model, this research used two main GLB policies, namely Power-aware and Cost-aware. The goals of these policies are to minimize the total power consumption and the total cloud electricity cost, respectively, summed across all geo-distributed data centers. Following that, this study used prediction results to offer geo-distributed data centers the possibility to participate in RS.

It is very difficult to predict power consumption of a data center that is part of a group of geo-distributed cloud data centers, especially with the intermittent availability of renewable energy and the use of free cooling devices like air economizers in data centers. The relationship between geo-distributed cloud data center power patterns and the weather characteristics (based on various conditions and infrastructure) is investigated in [31], and a set of influential features such as humidity, pressure, temperature, cloudiness, and wind speed, are extracted using linear regression (LR). The acquired features are then used to create an NN-based power consumption prediction model that predicts the power consumption pattern of each participating data center in a cloud.

The work in [32] presented an ML-based predictive resource allocation methodology for geo-distributed cloud data centers, which considered latency and quality constraints to ensure the best possible QoS for viewers and the lowest possible cost for live streaming content providers. To begin, this work proposed an offline optimization that determined the required transcoding resources in distributed locations near the viewers, balancing QoS and total cost. Next, ML is used to create forecasting models that anticipate the approximate transcoding resources that will be reserved at each cloud data center in advance. There are five different algorithms used namely, LSTM, GRU, CNN, MLP and XGBoost. The mean absolute error (MAE) metric is used to assess these regression models. Because there is no way to predict which combination of hyperparameters, such as the number of hidden layers and neurons for LSTM, GRU, CNN, and MLP models, or the number of estimators for XGBoost models, is optimal, the authors generated numerous models for each ML approach. The best models were chosen based on the best determination coefficient value, which is used to evaluate the regression models' goodness of fit. Finally, the authors created a Greedy Nearest and Cheapest Algorithm (GNCA) for allocating live incoming videos to the forecasted transcoding resources. The authors used Facebook video metadata as an input and considered 10 AWS cloud instances as a geo-distributed cloud platform for their analysis.

### 6.2.2 Workload Prediction

Because it is built for a single data center, the existing data analytics software stack is insensitive to on-the-fly resource variations over inter-data center networks, which can have a severe impact on query performance. To address this issue, [33] proposed Turbo, a light and non-intrusive data-driven framework that dynamically adapts query execution plans for geo-distributed data analytics in response to runtime resource fluctuations across data centers. Turbo uses two ML regression models, GBRT and LASSO, to precisely predict the time cost of query execution plans. The authors created a new dataset with 15K samples, each recording the time it took to run a query from the TPC-H benchmark [34], the size of the output, and a number of other parameters relevant to query execution. They set up a 33-instance cluster on the Google Cloud Compute Engine in eight different regions. Turbo is non-intrusive which means it does not necessitate any changes to the existing data analytics software stack.

Accurate workload forecasting is critical for dynamic resource scaling in geo-distributed cloud data centers. In [35], the authors proposed Savitzky-Golay and Wavelet-supported Stochastic Configuration Networks (SGW-SCN), an integrated forecasting approach with noise filtering and data frequency representation for predicting the workload in future time slots. The workload traces in Google production compute clusters from May 2011 were studied in this work. Over the course of 29 days, traces from a 12.5k-machine cell were collected, yielding a total of 25,462,157 tasks. Google clusters organize all tasks into multiple tiers, and each task has a characteristic reflecting its relevance. A higher level indicates that the task is more critical. Twelve attributes were investigated, which were divided into three types: gratis (0–1), other (2–8), and production (9–11). After that, the authors divided 29 days into 8352 five-minute time slots and counted the aggregated arrival rates of three different types of tasks. They used the data from the first 25 days for training and the last 4 days for testing. The workload time series is smoothed with a Savitzky-Golay filter before being split into several components using wavelet decomposition in this method. An integrated model is developed using SCN to characterize statistical properties of trends for different workload time series.

Data center availability is a crucial requirement in geo-distributed cloud environments. In this context, preventative efforts that reduce the time it takes to fix a data center's service in the event of a breakdown are critical. The authors in [36] proposed a method for determining the influence of network performance on service availability. This study considered six data centers around the world and calculated delay and jitter using Round Trip Time (RTT) and the download transfer rate in Mbps. Using ML algorithms capable of forecasting the time to transfer a large amount of data based on delay and jitter, this work developed smart agents capable of selecting the optimal data center to restore the services of a malfunctioning one. The authors selected LR, Poly LR, DT, RF, SVM, and MLP regression algorithms. They developed ten distinct train/test splits, assessed their accuracy with the Mean Absolute Error (MAE) metric, and then averaged the results.

### 6.2.3 Optimization Model Parameter Prediction

The authors of [37] investigated the topic of VM placement that reduces energy usage, $CO_2$ emissions, and access latency to reduce operating costs for geo-distributed cloud data center providers. This challenge is defined as a multi-objective function, with an intelligent ML model built to improve the presented Power and Cost-aware VM placement (PCVM) model's performance. The PCVM aims to reduce total costs by minimizing the weighted sum of two key objectives: carbon emission and network communication costs. While deciding among the two normalized model parameter weights associated with the two objectives, each one can be in conflict with one another. The authors applied a $k$-NN regression method to predict model parameter weights.

## 6.3 Optimization

We discuss prior work in the optimization category across two sub-categories: 'workload to resource mapping' and 'VNF placement'. Workload to resource mapping optimization refers to assigning incoming workloads (tasks or applications) to the best hardware resources across data centers. VNF placement optimization refers to dynamically placing VNFs on appropriate hardware to achieve an objective.

### 6.3.1 Workload to Resource Mapping Optimization

The authors of [38] proposed an adaptive resource management technique based on RL, with the goal of achieving a balance between QoS revenue and power usage across a set of geo-distributed cloud data centers. The proposed Q-learning based RL technique with custom random action selection is robust in allocating workload and does not require prior distribution of resource requirements. Instead of employing SLA violations, this work explicitly modeled QoS revenue using differential revenue of separate jobs. The authors considered the time spent migrating VMs across data centers and the cost of network communications. This work achieved quick decision-making by fine-tuning the RL algorithms' information storage and random action selection.

In a set of geo-distributed cloud data center environments, the cost of network resources is complicated to calculate because it is dependent on complex and heterogeneous inter data center WAN links. [39] abstracted geo-distributed data centers into an incomplete undirected graph. The problem of cloud data center selection is developed as the shortest-path problem from the client/origin vertex to all cloud data center vertices. The edge lengths in the graph were weighted, where the weights represent the distance between data centers or between the data center and a user. The authors considered the weights of the edge (network resource) and the vertex (computation resource) while selecting a data center. For mapping incoming workloads to an appropriate data center, they proposed a Q-learning based data center selection technique to achieve the optimal networking and compute costs. The state space comprised of a set of vertices of the graph and the action space is made up of the graph edges. After the workload moves from one vertex (state) along the edge (action) to another vertex (state), the length of the edge was used as the positive reward value. Through the Q-table, the path with the smallest Q value, i.e., the shortest path between each user and data center, was obtained.

In [40], the authors presented sCloud, a holistic heterogeneity-aware cloud resource management strategy, with the goal of maximizing system throughput in geo-distributed self-sustainable data centers. While taking into consideration renewable power availability and QoS requirements, sCloud adaptively distributed workloads to geo-distributed data centers. The proposed method assigned available resources to heterogeneous workloads in each data center and migrated batch jobs between data centers. The workload placement problem is formulated as a constrained optimization problem that can be solved using nonlinear programming. Moreover, when the green power supply fluctuated greatly at different places, the authors suggested a batch job migration strategy to boost system throughput even further. Finally, the authors enhanced sCloud by adding a Q-learning based RL technique. The job progress for batch jobs at separate data centers are represented by the elements in the RL state space. The action space is a collection of the progress control factors. The progress control factors determined how long the batch job can be delayed or accelerated compared to the job execution progress achieved in the previous control interval. The reward is the ratio of current goodput to a reference goodput plus penalties incurred when job level SLOs are violated. When tuning resource allocations, a low reward implies that the job execution may miss the soft or hard deadline, both of which should be avoided. The proposed RL based configurable batch job manager allowed dynamically regulation of the job execution process while staying on schedule.

Due to the dynamism of live broadcasting and the broad geo-distribution of viewers and broadcasters, satisfying all requests with adequate resources from geo-distributed data centers remains a challenge. To address this issue, [41] pre-sented a prediction-based approach that assessed the potential number of viewers at distinct cloud sites at the time of broadcasting. This work developed an Integer-Linear Program (ILP) based on the derived predictions to proactively and dynamically identify the best data centers to allocate exact resources and serve potential viewers while reducing perceived delays. Because ILP-based optimization is time consuming and thus inefficient for online serving, the authors presented RL-OPRA, a real-time DQN based RL technique. The authors defined a state as the set of data center index, predicted viewers, cost incurred from previous time-steps, and incremental average delay. An action indicated the index of the site that will serve the viewers. The total reward was determined by adhering to the average delay and reducing the cost of various actions. The proposed RL based solution adaptively learned to optimize allocations and service decisions while interacting with the network environment.

In [42], the problem of energy-aware workload management in geo-distributed cloud data centers considering their green energy generation with variable capacity was investigated. This work introduced a blockchain-based decentralized cloud management architecture to reduce the overall energy cost, workload scheduling cost, and workload migration in data centers. In this decentralized cloud management, the workload is scheduled by data centers themselves, removing the dependency on a central schedular. In addition, the authors proposed a Q-learning based RL technique incorporated in a smart contract to reduce energy cost even further. The state vector is comprised of data centers' execution load, the reward vector is made up of energy costs, and actions contain all workload migration possibilities. The proposed RL approach migrates workloads among geo-distributed data centers using the information about historical migration decisions and tries to minimize the total energy cost.

The data centers located in buildings that also provide significant space for office rooms are said to inhabit mixed-use buildings (MUBs). It is beneficial to use scheduling flexibility in both the heating, ventilation, and air conditioning (HVAC) system and the data center workload to successfully lower the total energy cost of MUBs. In [43], the authors proposed two DRL frameworks for lowering overall energy cost while maintaining target room temperature and satisfying data center workload deadline constraints. In case of the DRL based algorithm for office HVAC control, airflow rate controls are the actions, temperatures in the various office zones are the states, and the cooling energy cost is the reward. In case of the DRL based data center workload allocation, a server's activation (on/off) control is the action, execution rate of a server is the state, and data center cooling and compute energy cost is the reward. Then, the authors presented a joint office HVAC and data center workload control algorithm in a single MUB. They also created a heuristic DRL-based method to enable interactive workload distribution among geo-distributed MUBs, resulting in even more energy savings. Here the number of servers preferred for use in each MUB is the action, execution rate of an MUB is the state, and total energy cost for all MUBs is the reward.

In [44], the authors discussed the problem of big data analytics cost reduction in geo-distributed data centers powered by renewable energy sources with intermittent capacity. To address this problem, a DQN based RL technique for workload scheduling was proposed. The state vector was composed of current information about the CPU usage, free RAM size, free I/O bandwidth, weather, and electricity price. The action vector consists of the job locations (information about where to migrate the jobs). The reward is comprised of migration costs and energy costs. In addition, two strategies are created to improve the framework's performance. Random Pool Sampling (RPS) is proposed to retrain the NN using collected training data, and a unique Unidirectional Bridge Network (UBN) structure is devised to improve the training time even further by leveraging the historical information stored in the trained NN.

VMs virtualize hardware to allow several OS instances to operate simultaneously. On the other hand, Containers allow users to virtualize an OS so that many workloads can operate on the same machine. Containers also make it possible to bundle a program with all of its dependencies (such as code and libraries), allowing for faster startup, shutdown, and migration times than VMs. The default container placement technique in Kubernetes is unsuitable for a geographically distributed computing environment or dealing with the volatility of computing resources and workload. [45] proposed ge-kube (Geo-distributed and Elastic deployment of containers in Kubernetes), an orchestrated solution that leverages Kubernetes and adds self-adaptation and network-aware placement features. This study provided a two-step control loop in which the number of replicas of specific containers are dynamically controlled based on the application response time using a model-based RL technique, and containers are assigned to geo-distributed computing resources using a network-aware placement mechanism. For the RL technique, the state is made up of number of application instances, their CPU utilization, and their CPU resource limit. The action has vertical (add or remove CPU share) and horizontal (add or remove containers) scaling control. The reward is total cost, which is defined as the weighted sum of the performance penalty (paid after a deadline miss), resource utilization cost, and the adaptation cost. The study proposed an optimization problem formulation and a network-aware heuristic to address the placement issue, which explicitly took into consideration non-negligible network delays across geo-distributed computing resources to satisfy latency-sensitive workload's QoS needs. This work conducted an extensive series of experiments using a surrogate CPU-intensive application and a real application (i.e., Redis), demonstrating the benefits of combining elasticity and placement strategies, as well as the advantages of adopting network-aware placement solutions.

The authors of [46] investigated intra- and inter-data center energy optimization with both homogeneous and heterogeneous servers and workloads. They initially proposed an optimization model to reduce the combined server and network energy costs. The authors used the Lyapunov optimization model to turn the original problem into a well-studied queue stability problem to address the time-coupling restriction of carbon emission. Then, using the generalized Benders decomposition method, the authors developed a centralized solution. They also enhanced the model to handle heterogeneous workloads and data centers. At the start of every time slot, this technique found a feasible solution for energy optimization.

But it followed this method for the entire time slot/epoch, despite changes in the network environment (e.g., occurrence of faults). To address this issue, the authors developed a DQN based RL approach. The state is the status of a data center which includes number of active servers and the frequency of CPU. The action considers a set of data centers and controls which data center to choose. The reward is the total long term energy cost. The proposed DQN based RL method was able to explore and learn the dynamic nature of the network and make run-time decisions based on limited knowledge.

### 6.3.2 VNF Placement Optimization

Service function chaining (SFC) aims to establish VNFs in geo-distributed data centers and facilitates interconnect routing between them. DRL has recently been shown to be beneficial in the field of SFC. But current DRL-based algorithms, possess a large action space which causes poor convergence and scalability. Some studies have found a solution to this challenge by reformulating the SFC problem, which usually leads to low utilization and steep costs. To address this problem, [47] created a hybrid DRL-based system that separates VNF deployment and flow routing into separate modules. A DRL agent's sole responsibility is to learn the VNF deployment policy. Adaptive parameter noise, Wolpertinger policy, and prioritized experience replay are used to adjust the structure of the agent based on DDPG and to increase learning efficiency. A game-based module (GBM) is used to route the flow. A decentralized routing algorithm for the GBM is designed to address the scalability. The DRL agent altered the deployment policy based on the reward generated by the GBM during the learning process. The proposed method outperformed traditional DRL-based algorithms in terms of learning efficiency for two reasons. First, the suggested technique drastically shrank the DRL agent's action space. Second, traditional DRL-based algorithms are almost entirely model-free. The GBM, on the other hand, used model-based algorithms to optimize flow routing in the proposed method. The model-based information, such as the gradient, led to notable improvements in the algorithm's efficiency.

### 6.4 Hybrid ML Management

Most cloud management solutions use one problem-solving technique in their operational workflow. But some are required to achieve multiple goals at the same time, such as predicting the incoming workload and then allocating it to resources, or clustering a set of resources based on the current execution state and then provisioning them dynamically. These workflows can benefit from separate ML techniques for each objective. Hybrid ML methods have been developed to overcome drawbacks of standalone techniques by combining one or more ML techniques together. A few workflows that we consider in this survey are required to predict some resource or workload parameters before starting the dynamic optimization process. Accordingly, the hybrid ML methods that we discus here combine parameter prediction and cloud optimization.

Designing efficient scaling algorithms is difficult, particularly for geo-distributed VNF chains, where WAN traffic bandwidth costs and latencies are essential, yet difficult to account for when scaling decisions are made. To resolve this problem, [48] offered a deep learning-based methodology, analyzing intrinsic patterns of traffic volatility and efficient deployment techniques over time, with the goal of improved judgments enabled by in-depth learning from experiences. An LSTM is used for predicting future flow rates, and a DRL agent is used to make dynamic VNF chain placement decisions. To improve results, this study used an experience replay technique based on the actor–critic DRL algorithm. In comparison to previous representative algorithms, trace-driven simulation showed that the newly designed learning framework reacted swiftly to traffic dynamics online and achieved lower system costs with minimum offline training.

Web app developers previously had to store more data object copies in many geo-distributed data centers or send duplicate requests to multiple (e.g., closest) data centers to provide low request latency to users, both of which increased monetary cost. To address this issue, [49] proposed an RL based geo-distributed cloud storage system named GeoCol, with the goal of achieving low cost and latency. First, this new system included a request split mechanism and a storage planning method to achieve the best tradeoff between monetary cost and request latency. To enable parallel transmissions for a data object, the request split approach used the SARIMA technique to anticipate request latency as an input to an RL model to calculate the number of sub-requests and the data center assigned to each sub-request. Second, GeoCol employed another RL method to decide if it needs to store each data object and the storage type of each stored data object to help reduce the storage cost.

In [50], the authors studied ways to connect different renewable energy generators to geo-distributed data centers from various cloud providers in order to reduce carbon emissions, monetary costs, and service level objective (SLO) violations caused by renewable energy shortages. The authors tested multiple ML approaches for long-term forecast accuracy on renewable energy generation and demand, and chose SARIMA for the prediction. Based on the projected findings, the study presented a multi-agent RL (MARL) approach for each data center to determine how much renewable energy to request from each generator. This research also presented a Deadline Guaranteed Job Postponement approach (DGJP) for deferring the execution of non-essential jobs when renewable energy supplies are insufficient.

## 7 DISCUSSIONS
### 7.1 ML Technique Categorization

To handle the problem of reliable cloud management for geographically distributed data centers, several sub-problems must be addressed. We looked at a variety of methods and strategies presented for such reasons in the previous section. Most use one, but a few of them use multiple ML techniques in their operational flow. According to the information presented in Section 1.3, ML approaches contain any models that can be trained using data to perform supervised, unsupervised, or RL based tasks. In Table 1, we categorize the ML models used in all papers discussed in this survey into four different types. Figure 3 shows the percentage of ML model types considered by various articles. In Table 2, we provide a comparative summary of the investigated papers based on the various ML approaches utilized in the geo-distributed cloud management environment. The table also demonstrates why different ML models are used in various research articles.

TABLE 1. ML methods used in the survey

| ML Type | ML Algorithms |
|---|---|
| Clustering | *k*-means, Spectral |
| Regression | LR, Poly LR, LASSO, DT, RF, GBRT, XGBoost, *k*-NN, SVM, SARIMA |
| Neural Networks | SCN, MLP, CNN, LSTM, GRU |
| Reinforcement Learning | Model-based, Actor-critic, Q-learning, DQN, DDPG, DRL, MARL |

From Figure 3, by looking at the limited usage of clustering-based learning techniques (*k*-means and spectral), we can say that these are the least suitable for cloud management because they group workloads or resources into clusters, but at a scale and granularity that is not particularly beneficial for cloud management frameworks. The only time when unsupervised clustering can be advantageous is when the workload or resources change very frequently.

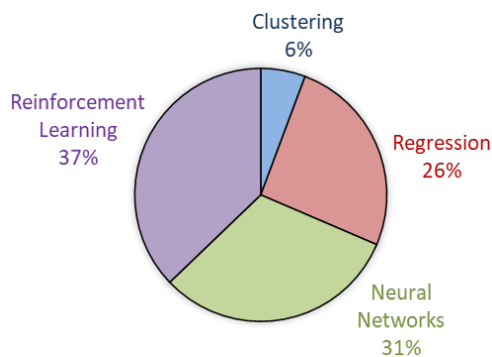

Fig. 3. Percentage of ML model types considered in prior work

As per the findings of this review, workload and resource parameter prediction is a critical task in large-scale geo-distributed cloud computing infrastructures. High-accuracy predictions enable reliable management of incoming workload and available resources, ensuring QoS and SLAs while lowering cloud operating costs. However, predicting workload and resource parameters is a difficult problem in general, especially when they are dependent on unpredictable activities (e.g., web servers, IoT devices, biological sensors, smartphone/PC workloads, resources failures). Regressive models, such as LR, Lasso, or SARIMA, are often used to make such predictions. A new generation of data analysis models based on NN, on the other hand, has just been embraced, and extensive evaluations of them have shown encouraging results. On average, RNN-based models, e.g., LSTMs, seem to produce better predictions quickly compared to traditional regression based models [32].

In practice, various challenging characteristics of geo-distributed cloud systems, e.g., massive scale and heterogeneity of systems, randomness of workload arrival parameters and resource availability, and timeliness of scheduling decisions, make it very difficult to design accurate models, which necessitates the use of a high-speed scheduling algorithm. It is difficult for a single Heuristic or Metaheuristic algorithm to adequately adapt to the real-world dynamics of geo-distributed cloud computing systems. In such scenarios, the profound use of RL algorithms in reviewed articles shows that it can be a very promising for cloud computing management. In the area of ML based optimization, RL is in fact the most prevalent method explored.

### 7.2 Challenges and Future Opportunities

Having discussed the current state-of-the-art in geo distributed cloud management, we now discuss several research challenges and opportunities in this section that address the open issues highlighted in current literature and hence represent viable future research directions in the area of ML for geo-distributed cloud data center management.

**Advanced ML algorithms:** Current studies use traditional ML algorithms such as regression, *k*-means clustering, Q-learning, etc. More advanced NN and Deep Learning (DL) techniques that have been shown to successfully and more accurately solve large scale problems in other application domains, could be adapted for the geo-distributed cloud management domain. For example, LSTMs can predict future values based on previous sequential data. They can be used where lightweight and fast time series forecasting models are needed. If a little more computational overhead is acceptable, Transformer ML models can be used instead of LSTMs. They can process much larger amounts of data in the same amount of time because of the parallelizability of the transformer mechanism. DL is especially useful when dealing with problems that have a high-dimensional state space. In DRL, because of DL's ability to learn different levels of abstraction from data, RL can perform increasingly sophisticated tasks with less prior information. Emerging Soft Actor-Critic (SAC) RL techniques can optimize a stochastic policy in an off-policy manner, bridging the gap between stochastic policy optimization and DDPG-style techniques. It employs entropy regularization, in which the policy is trained to maximize the trade-off between expected return and entropy (randomness in the policy). Due to entropy factor regularization, SAC can be highly efficient for energy-based optimization approaches.

**Efficient ML training and deployment:** One component of the design of parallel ML algorithms is to leverage several threads/processes to speed up the ML model's learning speed. In some cases, NNs require a large amount of training data, a significant amount of processing time, and even specific hardware to run parallel ML algorithms, such as servers with GPUs. Instead of using such computationally heavy ML algorithms (e.g., NN, Deep Learning, RL, etc.), researchers need to propose fast and lightweight ML solutions that can be deployed on commodity hardware. Model compression techniques (pruning, quantization, Low-rank factorization, Knowledge distillation, etc.) are particularly promising to reduce model inference overheads., particularly if the ML models are executed frequently. Model compression is a strategy for deploying cutting-edge ML models with lower memory usage and reduced latency, without sacrificing accuracy. Many cloud management studies have developed efficient ML-based profiling, prediction, or optimization algorithms. But most of these ML techniques are not designed to work online (for inference) and make real-time decisions after processing large volumes of real-time data. These existing techniques could be converted to online ones (e.g., RL or DRL based algorithms) to work with profiling and prediction methods, with model compression to improve efficiency. Another element to investigate is using ensemble learning and RL together for quick deployment.

**Benchmarks/Workloads:** Most proposed ML models have not been trained and evaluated using big, high-quality datasets obtained by strong industrial cloud operators in production scenarios. The majority of prior work uses synthetic, small datasets, or datasets that are not reflective of real-world geo-distributed events. Because there is a scarcity of current and relevant data (for workloads, resource usage, and trends) from geo-distributed cloud environments, it is difficult to accurately assess the quality of published results and, more importantly, to compare the results of competing studies. A large scale, realistic, publicly available geo-distributed cloud data center dataset from a major cloud vendor would serve as a standard, allowing researchers to analyze and compare alternative ideas and approaches on a much larger scale.

**State-of-the-art Distributed Computing Systems:** A logical step would be to explore the design and construction of a distributed version of the proposed ML algorithms and build the complete management system using cutting-edge big data technologies, e.g., Apache Spark, that can be deployed in geo-distributed cloud data centers.

**Industry Involvement for Validation:** It would be ideal to have more industrial participants involved in the studies, not only by supplying appropriate requirements, but also by contributing to the evaluation of the results, and especially the deployment of small-scale pilots in production settings. Although small testbeds that are used in some studies allow for the testing of many ideas and configurations, it is impossible to draw definitive or full findings without a thorough examination in a production setting.

**Heterogeneous Resources/Hardware:** Most studies consider traditional hardware resources e.g., server nodes, CPUs, etc. The proposed approaches in these studies could be extended to consider more complicated user requests with multiple types of resources other than CPUs, e.g., disaggregated memories, GPUs, accelerators, network switches, emerging cooling systems, and considering heterogeneity between them.

**Networks:** Inter and intra-data center networks play a very important role during cloud management (reducing operating costs, QoS, SLA, etc.). With ever increasing data volumes in new cloud workloads, it is very important to factor in inter as well as intra data center networks, and the network costs related to the data transfers across geo-distributed datacenters [25]. Studies must include network related parameters in the optimization function to resemble state-of-the-art cloud management systems.

**Energy Modeling:** Most if not all studies included in this survey consider simple data center energy models that are based on CPU/node utilization. In the data center energy model, for improved accuracy, researchers could consider various factors such as different performance states (P states) of cores, thermal aware cooling power, net metering and peak shaving, i.e., peak power reduction, etc. [24].

***Multi-objective Algorithms:*** Unlike many existing studies, real world cloud management systems must address more than one or two objectives. Objectives of future studies can be modified to include more optimization metrics. For example, data center computing/network performance (latency, SLAs, QoS, etc.) and data center operating/network costs could be optimized jointly; data center brown energy consumption, renewable energy generation, and carbon emission could be considered; and data center performance, uptime/availability, and fault tolerance could also be optimized jointly.

## 8 CONCLUSIONS

In this survey, we reviewed how the challenges of reliable geo-distributed cloud management has been addressed using ML techniques in the scientific literature. For simplification, we dissected this problem into three sub-categories: profiling, parameter prediction, and cloud optimization. Moreover, we discussed how ML methods have been used to improve the reliability of the cloud management algorithms in diverse and heterogeneous environments. The number of studies that use ML based approaches in this area has increased dramatically in recent years. Several researchers have employed a variety of methodologies, ranging from traditional statistics and regression to advanced ML algorithms. We conclude that, on average, ML methods outperform traditional mathematical techniques, particularly when dealing with large and complicated environments. The progression of ML methodologies in current research is illustrated in this article, which we hope will help readers comprehend the research gaps in this topic. Finally, based on the challenges discovered in the surveyed studies, several prospective future research directions and opportunities are identified to strengthen the current ML approaches.

## ACKNOWLEDGMENTS

This work is supported by the National Science Foundation (NSF) under grant CCF-1302693.

TABLE 2: Summary of Machine Learning Techniques Used in Geo-distributed Cloud Data Center Management

| ML Type | Ref. | ML Functionality | ML Model | Main Results |
|---|---|---|---|---|
| resource profiling | [26] | cluster data centers according to four SLA attributes - cost, response time, availability, and reliability | $k$-means clustering | selected the best data center according to SLA rules and the user's request |
| workload profiling | [28] | partition a set of data-items into geo-distributed clouds | spectral clustering | performed up to 10 times (3–4 times on average) faster |
| resource prediction | [30] | predict data center incoming workload locally and power consumption globally | SARIMA & CNN | reduced the energy cost by 22% on average |
| resource prediction | [31] | predict the power pattern | LR & NN | power prediction model reached the accuracy of 87.2% |
| resource prediction | [32] | predict the approximate transcoding resources to be reserved | GRU, LSTM, MLP, and CNN, and XGBoost | achieved more than 20% gain in terms of system cost |
| workload prediction | [33] | accurately estimate the time cost of query execution plans | LASSO Regression and GBRT | achieved cost estimation accuracy of over 95% and reduce query completion times by 41% |
| workload prediction | [35] | predict the workload in future time series | SGW-SCN | produced more accurate forecasting results and with faster learning speed |
| workload prediction | [36] | predict delay, jitter, and throughput of each network link | LR, Poly LR, DT, RF, SVM, MLP | chose the best data center to restore the services of a faulty one |
| model parameter prediction | [37] | predict optimization model weights | $k$-NN | improved cloud provider net profit by reducing DCs power consumption |
| workload to resource mapping optimization | [38] | get the balance between QoS revenue and power consumption | RL (Q-learning) | power consumption is up to 13.3% better |
| workload to resource mapping optimization | [39] | search for the shortest path while minimizing network costs | RL (Q-learning) | effectively solved the problem achieving low network and compute resource costs |
| workload to resource mapping optimization | [40] | allocate resources to workloads while considering the green power availability and QoS requirements | RL (Q-learning) | improved 37% system goodput and saved 33% QoS violations |
| workload to resource mapping optimization | [41] | choose the right data center to allocate resources and serve potential viewers (for live streaming), while minimizing delays and operating costs | RL (DQN) | outperformed comparison heuristics while adapting to the dynamics of the crowdsourcing system, through continuous learning |
| workload to resource mapping optimization | [42] | maximize cloud operating profits | RL (Q-learning) | saved up to 50% more energy costs |
| workload to resource mapping optimization | [43] | schedule HVAC control and workload for minimizing the total energy cost while maintaining desired temperature and meeting workload deadlines | RL (DRL) | achieved up to 26.9% energy cost reduction for a single location, and an additional 5.5% for all data centers combined |
| workload to resource mapping optimization | [44] | minimize costs of big data analytics on data centers connected to renewable energy sources with unpredictable capacity | RL (DQN) | reduced the data centers' energy and data migration cost significantly |
| workload to resource mapping optimization | [45] | control the number of container replicas based on the application response time, and allocate containers with the help of a network-aware placement policy | RL (model-based) | processed 2.91 times more operations per second |
| workload to resource mapping optimization | [46] | joint energy consumption optimization of servers, and the network | RL (DQN) | has high efficiency in cost-saving and performance-enhancing |
| VNF optimization | [47] | learn the policy of VNF deployment | DRL (DDPG) | average network savings are up to 29.8% |
| workload prediction & resource optimization | [48] | predict upcoming flow rates; then make chain placement decisions | LSTM; DRL (actor-critic) | reduced total system cost by up to 42% |
| workload prediction & resource optimization | [49] | predict request latency and split across data centers; then use storage planning to reduce monetary cost | SARIMA & RL; RL | achieved 32% monetary cost reduction and 51% data object request latency reduction |
| resource prediction & workload optimization | [50] | predict the amount renewable energy generation and its demand at each data center; then determine how much renewable energy to request from each generator | SARIMA; MARL | increased up to 35% SLO satisfaction ratio, and reduced up to 19% (0.33 billion dollars in 90 days) total monetary cost and 33% total carbon emission |

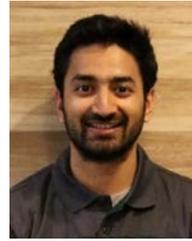

**Ninad Hogade** received his B.E. degree in Electronics Engineering from Vishwakarma Institute of Technology, India, and M.S. degree in Computer Engineering from Colorado State University, USA. He is currently a Ph.D. student in Computer Engineering at Colorado State University, USA. His research interests include energy aware scheduling of high performance computing systems and data centers.

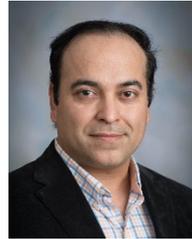

**Sudeep Pasricha** received his B.E. degree in Electronics and Communications from Delhi Institute of Technology, India; and his M.S. and Ph.D. degrees in Computer Science from University of California, Irvine. He is currently a Professor and Chair of Computer Engineering at Colorado State University, where he is also a Professor of Computer Science and Systems Engineering. He is a Senior Member of the IEEE and an ACM Distinguished Member. Homepage: http://www.engr.colostate.edu/sudeep.